\documentstyle[pre,aps,epsfig,multicol,amssymb]{revtex}
\begin{document}
\draft

\title{Mean Field Fluid Behavior of the  Gaussian Core Model}
\author{A.A. Louis, P.G. Bolhuis, and
J.P. Hansen} \address{Department of Chemistry, Lensfield Rd,
Cambridge CB2 1EW, UK\\} \date{\today} \maketitle
\begin{abstract}
We show that the Gaussian core model of particles interacting via a
penetrable repulsive Gaussian potential, first considered by
Stillinger (J.\ Chem.\ Phys.\ {\bf 65}, 3968 (1976)), behaves like a
weakly correlated ``mean field fluid'' over a surprisingly wide
density and temperature range.  In the bulk the structure of the fluid
phase is accurately described by the random phase approximation for
the direct correlation function, and by the more sophisticated HNC
integral equation.  The resulting pressure deviates very little from a
simple, mean-field like, quadratic form in the density, while the low
density virial expansion turns out to have an extremely small radius
of convergence.  Density profiles near a hard wall are also very
accurately described by the corresponding mean-field free-energy
functional.  The binary version of the model exhibits a spinodal
instability against de-mixing at high densities.  Possible implications
for semi-dilute polymer solutions are discussed.
\noindent 

\end{abstract}
%61.20.-p Structure of liquids
%61.25.Hq Macromolecular and polymer solutions; polymer melts; swelling
%83.70.Hq Heterogeneous liquids: suspensions, dispersions, emulsions, pastes, slurries, foams, block copolymers, etc.

\begin{multicols}{2}

\section{Intro}

\begin{figure}
\begin{center}
\epsfig{figure=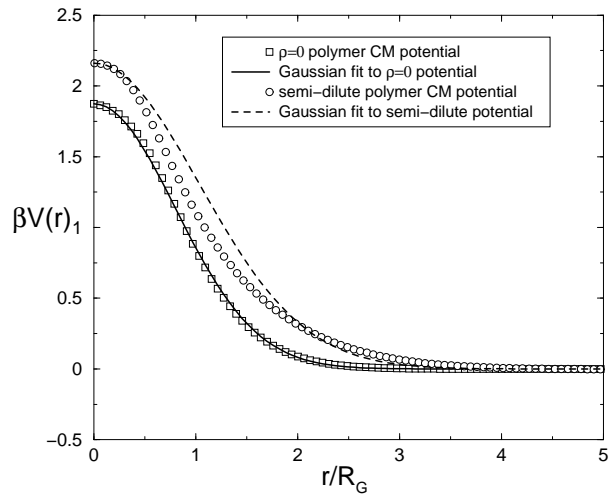,width=8cm}
\begin{minipage}{8cm}
\caption{\label{Fig1.1} Polymer center of mass potentials $\beta v(r)$ from
simulations of $L=500$ monomer SAW chains\protect\cite{Loui00a} are compared to
a best-fit Gaussian (\protect\ref{eq1.1}), determined by fitting
$\beta v(0)$
to fix $\beta \epsilon$, and  $\beta \hat{v}(0)$ to fix $R$.  The
 potential for two isolated coils ($\rho \rightarrow 0$)
 is well approximated by a Gaussian
potential with $\beta \epsilon = 1.87, R=1.13R_G$.  The potential in the
semi-dilute regime ($\rho \sim 4 \times 3/(4\pi R_g^3) $) is approximated by
a Gaussian potential with $\beta \epsilon = 2.16, R=1.45R_G$.  }
\end{minipage}
\end{center}
\end{figure}

Interactions between atoms or molecules in simple fluids invariably
contain a short-range repulsive component or hard core, such that the
local molecular structure is dominated by excluded volume effects.
This observation explains the success of simple models involving hard
convex bodies in explaining the structure and phase transitions in
simple atomic or molecular fluids\cite{Chan83}.  For example, the hard
sphere model has been instrumental in understanding freezing of simple
fluids\cite{1}.  The same extends success to somewhat more complex
fluids like liquid crystals, where hard ellipsoids or spherocylinders
have been widely used to investigate the isotropic-to-nematic
transition and other mesophases\cite{2}.  However the situation is
generally not as simple in complex fluids, where interactions between
mesoscopic particles are often of effective and of 
 entropic
origin.  While excluded volume effects still dominate the interaction
between compact colloidal particles, the effective forces between
``soft'' or fractal objects of fluctuating shape, like polymer coils
or membranes, cannot be modeled by hard cores.  Polymers in a good
solvent form highly penetrable coils and it is by now well established
that the effective interaction between the centers of mass of two
polymer coils, duly averaged over internal conformations, is finite
for all distances, and decays rapidly beyond the radius of gyration of
the coils\cite{Gros81,Krug89,Daut94}.  For two isolated
non-intersecting polymer chains, the effective pair potential at zero
separation of the centers of mass $v(r$$=$$0)$, is of the order of $2
k_B T$ for sufficiently long chains\cite{Krug89,Daut94}, and is
reasonably well represented by a Gaussian whose  width is of order the
polymer radius of gyration $R_G$, as shown in Fig.\ \ref{Fig1.1}.

  We have recently shown that the general shape of
the effective pair potential remains roughly the same in dilute and
semi-dilute solutions of self-avoiding random walk (SAW) polymers, and
does not vary strongly with polymer concentration (cf.\ Fig.
\ref{Fig1.1})\cite{Loui00a}. The effective pair potential model has
been shown to accurately reproduce the structure and thermodynamics
calculated from Monte Carlo (MC) simulations of solutions of SAW
polymers over a wide range of concentrations\cite{Loui00a}.

Neglecting in the first instance the state dependence of the effective
potential, it seems hence worthwhile to examine the equilibrium
properties of a fluid of ``soft'' particles interacting via a pair
potential approximated by a simple Gaussian form:
\begin{equation}\label{eq1.1}
v(r) = \epsilon \exp\left(- \frac{r^2}{R^2} \right)
\end{equation}
where $\epsilon$ is the energy scale, and $R$ determines the width.
The Fourier transform (FT) is
\begin{equation}\label{eq1.2}
\hat{v}(k) = \pi^{3/2}R^3\epsilon \exp\left(- \frac{k^2 R^2}{4} \right).
\end{equation}
Such a ``Gaussian core model'' (GCM) was in fact introduced some time ago by
Stillinger\cite{Stil76}, who focussed on the low temperature regime
$ \epsilon^* = \epsilon/k_B T >> 1$, where the model exhibits hard-sphere
like behavior, and a re-entrant fluid-solid-fluid phase diagram under
compression below a threshold temperature.  This work was further
expanded by Likos {\em et al.}\cite{Lang00}, who showed that the model
remains fluid at all densities when $\epsilon^* \lesssim 100$.  They also
demonstrated that for this model, the familiar HNC closure for the
pair distribution function $g(r)$ becomes exact in the high density
limit, and that the random phase approximation (RPA) is remarkably
accurate at high densities.

In this paper we concentrate on the fluid phase of the GCM ($\epsilon^*
<100$), with a particular emphasis on the regime relevant for polymer
solutions ($\epsilon^* \simeq 2$)\cite{Loui00a}, for which the dilute regime
corresponds to reduced densities $\rho^* = \rho R^3 \lesssim
3/(4\pi)\approx 0.239$, and the semi-dilute regime corresponds to
$\rho^* \gtrsim 3/(4\pi)$\cite{polymer} (here $\rho = N/V$ is the
number of Gaussian core particles per unit volume).  We shall
successively consider the homogeneous fluid phase, the inhomogeneous
fluid phase in the vicinity of a hard wall, and the possibility of
de-mixing of binary Gaussian core systems.

\section{The homogeneous fluid phase}\label{secII}

\subsection{The Thermodynamic stability of the GCM fluid}\label{secIIa}

We consider a system of N particles interacting via a Gaussian pair
potential (\ref{eq1.1}), in a volume $V$.  In the absence of an
infinitely repulsive core the first question is that of thermodynamic
stability against collapse, i.e.\ the existence of a well defined
thermodynamic limit.  According to definition 3.2.1.\ in Ruelle's
classic book\cite{Ruel69}, the total interaction energy $V_N$, which can be
built up of pair and higher order potentials, is {\em stable} if there
exists a $B \geq 0$ such that
\begin{equation}\label{eq2.2}
V_N({\bf r_1}, ....,{\bf r_N}) \geq -N B
\end{equation}
for all $N > 0$ and all $\{{\bf r_i}\}$ in the phase-space $R^N$.
Stability implies convergence of the grand partition function and a
well defined thermodynamic limit.  Specializing to pair potentials
$v_2$, the total potential energy of the system, for any configuration
of N particles $\{{\bf r_i}\} \in R^N$, can be written as :
\begin{equation}\label{eq2.1}
V_N^{(2)}({\bf r_1}, ....,{\bf r_N}) = \sum_{1 \leq i < j \leq N} v_2(|{\bf
r_i} - {\bf r_j}|)
\end{equation}
For purely repulsive pair potentials, such as the GCM with $\epsilon^* \geq 0$,
$V^{(2)}_N$ satisfies the condition (\ref{eq2.2}), so that a well defined
thermodynamic limit exists.  However, if $v_2(r)$ is not strictly
positive, this may no longer be true.  In Appendix A two
examples are discussed, involving a finite core and (small) attractive
tail, which do not lead to a proper thermodynamic limit.

\subsection{The Structure of the GCM fluid}

\begin{figure}
\begin{center}
\epsfig{figure=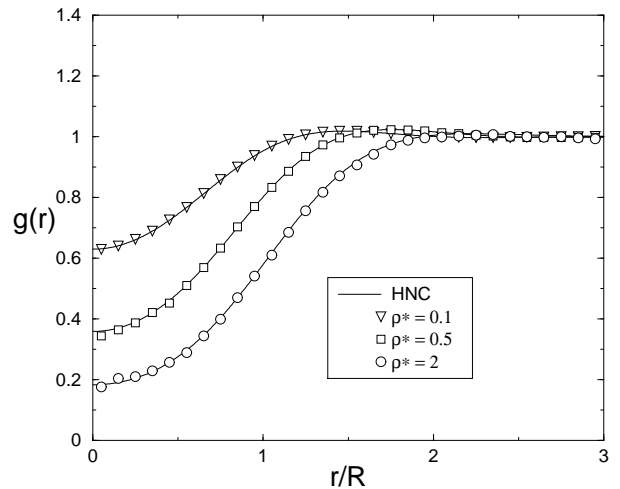,width=8cm} 
\begin{minipage}{8cm}
\caption{\label{Fig2.1} Comparison of MC simulations and solutions of
the HNC integral equation in a
regime relevant for polymer solutions\protect\cite{Loui00a}, 
 $v(r) = 2\exp[-(r/R)^2]$.
The lines are HNC calculations, and the symbols represent MC
simulations  for different reduced 
densities $\rho^*$. }
\end{minipage}
\end{center}
\end{figure}

\begin{figure}
\begin{center}
\epsfig{figure=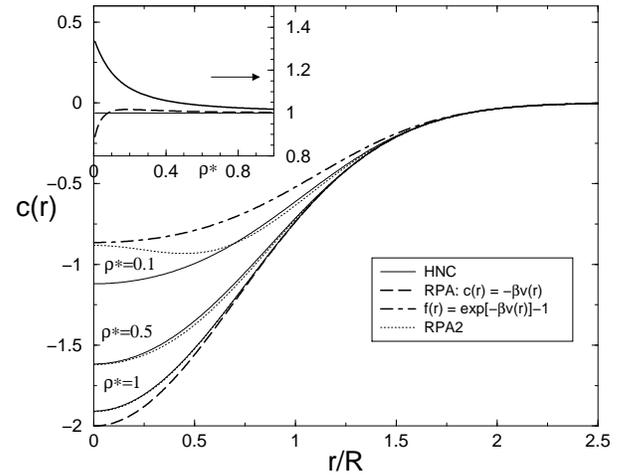,width=8cm} 
\begin{minipage}{8cm}
\caption{\label{Fig2.2} HNC, RPA, and RPA2 forms of the direct
correlation function $c(r)$ for $v(r) = 2 \exp[-(r/R)^2]$.  From top
to bottom the densities are $\rho^* = 0.1,0.5$, and $1$ respectively.  Note
that the HNC $c(r)$  is bounded by $c_{RPA}(r) = -\beta v(r)$ from below and
$f(r) = \exp[-\beta v(r)]-1$ from above.  {\bf Inset}: Ratio's
$\hat{c}_{RPA}(0)/\hat{c}_{HNC}(0)$ (solid line) and
$\hat{c}_{RPA2}(0)/\hat{c}_{HNC}(0)$ (long-dashed line) v.s.\ density
$\rho*$.  For $\rho* > 0.05$, $\hat{c}_{RPA2}(0)$ is always within
$2\%$ of $\hat{c}_{HNC}(0)$ }
\end{minipage}
\end{center}
\end{figure}

\begin{figure}
\begin{center}
\epsfig{figure=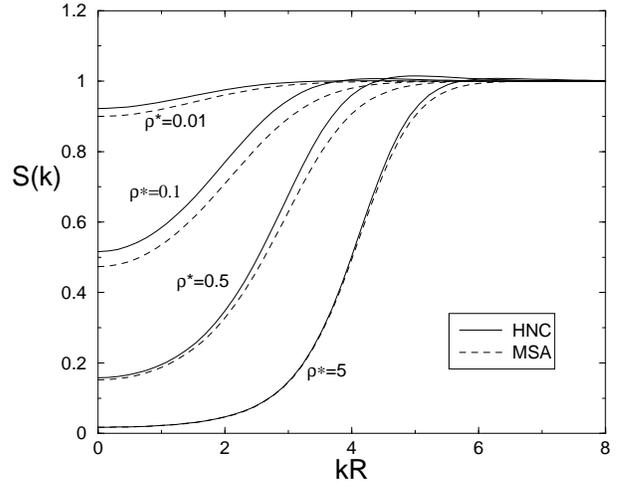,width=8cm} 
\begin{minipage}{8cm}
\caption{\label{Fig2.3} RPA and HNC forms of the structure factor
$S(k)$ for $v(r) = 2 \exp [-(r/R)^2]$.  From top to bottom the
densities are $\rho* = 0.01, 0.1, 0.5, 5$ respectively.}
\end{minipage}
\end{center}
\end{figure}

To determine the pair structure of the GCM fluid, we have used the
HNC closure which becomes exact in the high density limit; this
closure relates the direct correlation function $c(r)$ to the pair
potential $v(r)$ and the pair correlation function $h(r) = g(r)-1$, according
to:
\begin{equation}\label{eq2.3}
c(r) = -\beta v(r) + h(r) -\ln \left[1 + h(r) \right],
\end{equation}
where $\beta = 1/k_B T$.  This closure must be combined with the
Ornstein Zernicke (OZ) relation between $c(r)$ and $h(r)$\cite{Hans86}
to yield a non-linear integral equation, which must be solved
numerically.  Examples for $\epsilon^*=3$ at three reduced densities
$\rho*$ are shown in Fig \ref{Fig2.1}, and compared to the results of
MC simulations.
\vglue - 0.5cm The key feature is that the ``soft'' correlation hole
 is gradually reduced as $\rho^*$ increases, a behavior typical of
 finite core potentials, which leads to overlap and ideal-gas like
 behavior of $g(r)$ in the high density limit.  Note that the HNC
 results are indistinguishable from the MC data, so that for
 $\epsilon^* \simeq 2$ the HNC correlation function will henceforth be
 considered as providing an ``exact'' reference to gauge simpler
 theories.  The simplest is the RPA\cite{Hans86,RPA}, which may be
 formally derived from the HNC closure (\ref{eq2.3}) by linearizing
 the logarithm, leading to:
\begin{equation}\label{eq2.4}
c(r) = -\beta v(r).
\end{equation}
Since Fig.\ \ref{Fig2.1} clearly shows that the amplitude of $h(r)$ is
rather small at high densities, we may expect the RPA closure to
become  more accurate as the density increases. For the
GCM, Eq.\ (\ref{eq2.4}) and Eq.\ (\ref{eq1.2}) imply the following FT of
$c(r)$:
\begin{equation}\label{eq2.5}
\hat{c}(k) = -\epsilon^* \pi^{3/2} R^3 \exp\left[ \frac{-k^2 R^2}{4}\right]
\end{equation}
and the OZ relation immediately yields the following RPA structure
factor:
\begin{eqnarray}\label{eq2.6}
S(k)&=& 1 + \rho \hat{h}(k) = \frac{1}{1-\rho \hat{c}(k)} \nonumber \\
 &=& \frac{1}{1 + \alpha \exp \left[ -k^2 R^2/4\right]},
\end{eqnarray}
where we have introduced the dimensionless coupling parameter:
\begin{equation}\label{eq2.7}
\alpha = \pi^{3/2} \beta \epsilon \rho R^3 = \pi^{3/2} \rho^* \epsilon^*
\end{equation}
HNC results for $c(r)$ and $S(k)$ at several densities are compared to
the RPA predictions in Figs.\ \ref{Fig2.2} and \ref{Fig2.3}.

 Since $h(r) \geq \ln[1 + h(r)]$, the HNC direct
correlation functions are bounded below by the RPA form
(\ref{eq2.4}).  Fig.\ \ref{Fig2.2} also shows that the HNC
$c(r)$ appears to be bounded above by the low density approximation:
\begin{equation}\label{eq2.8}
c(r) = f(r) = \exp\left[- \beta v(r) \right] -1
\end{equation}
which corresponds to the lowest order term in the expansion of $c(r)$
in powers of $\rho$\cite{Hans86}; $f(r)$ is the usual Mayer
$f$-function.  Figs.\ \ref{Fig2.2} and \ref{Fig2.3} also illustrate the
point that the simple RPA becomes very accurate at high
densities, so that it is worthwhile to inquire about a correction to
Eq.\ (\ref{eq2.4}). Expanding the logarithm on the r.h.s.\ of
Eq.\ (\ref{eq2.3}) to second order in $h$, one arrives at the following
expression for $c(r)$:
\begin{equation}\label{eq2.9}
c(r) = -\beta v(r) + \frac{1}{2} h(r)^2.
\end{equation}
Solution of the closure (\ref{eq2.9}) and the corresponding OZ
relation requires an iterative procedure, as for the full HNC closure.
Further simplification amounts to replacing $h(r)$ in Eq.\ (\ref{eq2.9})
by its RPA form derived from Eq.\ (\ref{eq2.6}) by FT; we refer to this 
non-iterative approximation as RPA2\cite{Abe}:
\begin{equation}\label{eq2.9b}
c(r) = -\beta v(r) + \frac{1}{2} h_{RPA}(r)^2.
\end{equation}
From Fig. \ref{Fig2.2} it is clear that  $c_{RPA2}(r)$ is
indistinguishable from the HNC results except at low densities
($\rho^* \lesssim 0.2$).  The limitations of RPA theory at low
densities become apparent by considering the resulting behavior of 
$g(r)$ at short distance. 

\vglue -0.5cm
\vglue - 0.5cm
 The zero separation value is easily derived
from the $r \rightarrow 0$ limit of the FT of Eq.\ (\ref{eq2.6}), with
the result:
\begin{equation}\label{eq2.10}
g_{RPA}(0) = 1 + \frac{\epsilon^*}{\alpha }  Li_{3/2}(-\alpha),
\end{equation}
where the nth polylogarithm is defined by:
\begin{equation}\label{eq2.11}
Li_{n}(x) = \sum_{k=1}^\infty \frac{x^k}{k^n}.
\end{equation}
for $|x| \leq 1$.  If $\epsilon^* < 1$, $g_{RPA}(0)$ is positive for
all densities $\rho^*$.  However, when $\epsilon^* > 1$ there is
always a reduced density $\rho^*_0$ {\em below} which $g_{RPA}(0) <
0$, which is unphysical.  For example, if $\epsilon^* = 2$,
$g_{RPA}(0) < 0$, for $\rho^* < \rho^*_0 = 0.3617$.  However, even for
$\rho^* < \rho^*_0$, the structure factor $S(k)$ is still reasonably
well described by the RPA because the deficiencies of $g(r)$ at small
$r$ do not strongly affect $S(k)$.  This is also illustrated in the
inset of Fig.\ \ref{Fig2.2}, where $\hat{c}_{RPA2}(0)$ is seen
approximate the quasi-exact HNC result to within $2 \%$ for densities
$\rho^* \gtrsim 0.05$, which is a significantly lower bound than
$\rho_0^* = 0.3671$.

\subsection{Thermodynamics of the GCM fluid}

Turning now to thermodynamic properties, the equation of state can 
be calculated via either of two routes\cite{Hans86}:
from the compressibility equation:
\begin{equation}\label{eq2.12}
\beta P = \int_0^\rho \frac{\partial \beta P(\rho')}{\partial \rho'} d\rho'
 = \int_0^\rho  \left[1 - \rho' \hat{c}(k=0;\rho')\right] d\rho'
\end{equation} 
where P denotes the pressure. For soft-core potential systems
the virial equation leads to:
\begin{equation}\label{eq2.13}
\beta P = \rho + \frac{1}{2} \rho^2 \hat{v}(k=0)  - 
 \frac{2 \pi}{3}  \rho^2 \int_0^\infty r^3
 \frac{\partial \beta v(r)}{\partial r} h(r) dr.
\end{equation}
\vglue -0.5cm
\begin{figure}
\begin{center}
\epsfig{figure=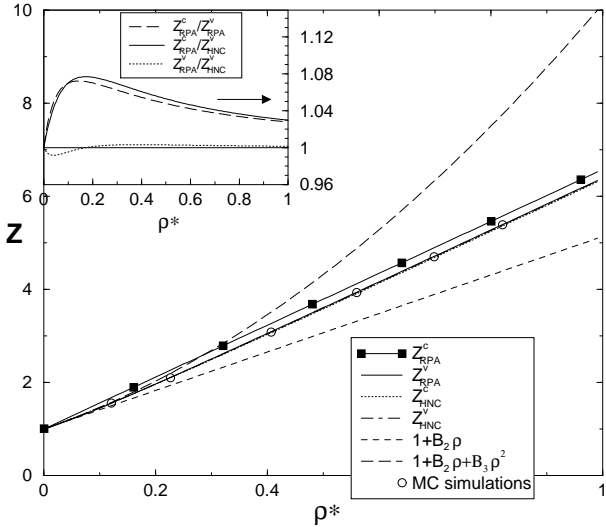,width=8cm} 
\begin{minipage}{8cm}
\caption{\label{Fig2.4} Compressibility factors ($Z = \beta P/\rho$)
from RPA and HNC for a Gaussian potential with $\epsilon^* =2$,
compared to MC simulations and to two and three term virial
expansions.  $Z_{RPA}^v, Z_{HNC}^c$, and $Z_{HNC}^v$, are
indistinguishable on this scale.  {\bf Inset:} The analytic ratio
$Z_{RPA}^c/Z_{RPA}^v= (1-\epsilon^* \aleph(\alpha)/(1 +
\alpha/2))^{-1}$ gives a good approximation to $Z_{RPA}^c/Z_{HNC}^v$
and goes to zero as $\rho \rightarrow \infty$.  The ratio
$Z_{RPA}^v/Z_{HNC}^v$ demonstrates that $Z_{RPA}^v$ approximates the true
e.o.s.\  to better than $1\%$ accuracy over the entire density range
for $\epsilon^*=2$.  }
\end{minipage}
\end{center}
\end{figure}
\vglue - 0.5cm

\begin{figure}
\begin{center}
\epsfig{figure=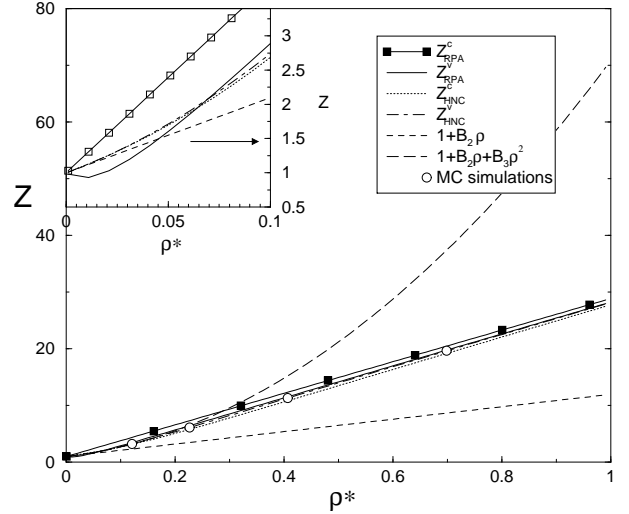,width=8cm} 
\begin{minipage}{8cm}
\caption{\label{Fig2.5}Compressibility factors 
from RPA and HNC for a Gaussian potential with $\epsilon^* =10$,
compared to MC simulations and to two and three term virial
expansions.
 $Z_{RPA}^v$,
$Z_{HNC}^c$, and $Z_{HNC}^v$ are very close over much of the density
range.
 {\bf Inset:}
Compressibility factors at low density, symbols are the same as in the
main figure.  Note that $Z^v_{RPA}$ shows
unphysical behavior for very small $\rho^*$, which can be understood 
from the effective virial expansion discussed in Appendix B.  }
\end{minipage}
\end{center}
\end{figure}

\begin{figure}
\begin{center}
\epsfig{figure=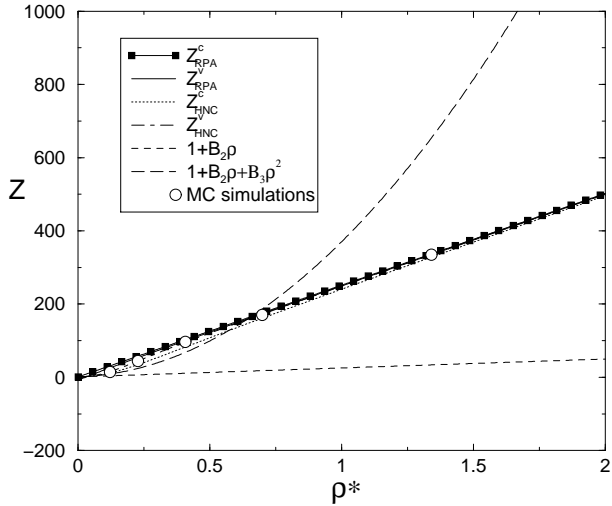,width=8cm} 
\begin{minipage}{8cm}
\caption{\label{Fig2.6}Compressibility factors from RPA and HNC for a
Gaussian potential with $\epsilon^* =90$, compared to MC simulations
and to two and three term virial expansions.  Again, the HNC virial
approximation is nearly exact across the whole density range, and the
e.o.s.\ is that of a mean-field fluid at all but the lowest densities.
The low-density limit is further discussed in Appendix B and illustrated 
in Fig.\ \protect\ref{FigB.3}.  }
\end{minipage}
\end{center}
\end{figure}

If the correlation functions $h(r)$ and $c(r)$ were known exactly, the
two  routes would lead to identical equations of state.   Approximate
theories are not, in general, thermodynamically consistent.  However,
as shown in Figs.\ \ref{Fig2.4}-\ref{Fig2.6}, the HNC closure yields 
practically identical values of the pressure over the whole range of
densities, even for a repulsion  as high as $\epsilon^* = 90$
(remember that for $\epsilon^* \gtrsim 100$ re-entrant crystallization sets
 in\cite{Stil76,Lang00}).  Moreover, the two HNC estimates of the
pressure 
agree closely with the results of MC simulations.  These results
confirm the conjecture that HNC and RPA become exact for the GCM in the high
density
limit, but also show that the HNC works well for low densities in the
fluid regime we consider ($\epsilon^* < 100$).

Turning now to the much simpler RPA, it is easily verified from 
Equations (\ref{eq2.4}) and
 (\ref{eq2.12}) that the dimensionless equation of 
state (e.o.s.), $Z = \beta P /\rho$, reduces, within the compressibility 
route, to the simple expression:
\begin{equation}\label{eq2.14}
Z_{RPA}^c = 1 + \frac{1}{2} \rho \beta \hat{v} (k=0)
\end{equation}
which for the GCM leads to:
\begin{equation}\label{eq2.14a}
Z_{RPA}^c =  1 + \frac{1}{2} \alpha. \label{eq2.14b}
\end{equation}
This in turn leads to an excess free energy per particle:
\begin{equation}\label{eq2.15}
\frac{\beta F^{ex}}{N} = \frac{1}{2} \rho \beta \hat{v}(k=0) =
\frac{1}{2} \alpha
\end{equation}
identical to that obtained from a van der Waals like mean field theory
(MF) so that $Z_{RPA}^c=Z_{MF}$; it also implies that the excess
chemical potential is linear in density.
\vglue - 0.5cm

Remembering that the quasi-exact HNC direct correlation function is
bounded below by the RPA form (\ref{eq2.4}), it is immediately clear
from the compressibility equation (\ref{eq2.12}) that one may expect:
\begin{equation}\label{eq2.16}
Z < Z_{RPA}^c = Z_{MF}
\end{equation}
This conjecture is supported by Eq.\ (\ref{eq2.13}), which shows that
the exact equation of state is given by:
\begin{equation}\label{eq2.17}
Z = Z_{RPA}^c -  \frac{2 \pi}{3}  \rho^2 \int_0^\infty r^3
 \frac{\partial \beta v(r)}{\partial r} h(r) dr,
\end{equation}
which, for the Gaussian core model reduces to:
\begin{equation}\label{eq2.18}
Z = 1 + \frac{1}{2} \alpha + \frac{4 \alpha}{3 \sqrt{\pi}}
 \int_0^\infty x^4 e^{-x^2} h(x) dx
\end{equation}
where the dimensionless spacing $x=r/R$ was introduced.  The conjecture
(\ref{eq2.16}) is thus true, provided the integral on the r.h.s.\ of
Eq.\ (\ref{eq2.18}) is negative.  This is very likely for sufficiently
high temperatures since the HNC results plotted in Fig.\ \ref{Fig2.1}
show that $h(x)$ is mostly negative.

The integral in Eq.\ (\ref{eq2.18}) can be calculated analytically
within  the RPA, leading to the following result for the RPA virial
equation of state:
\begin{equation}\label{eq2.19}
Z_{RPA}^v = 1 + \frac{1}{2} \alpha -  \epsilon^*\aleph(\alpha) 
\end{equation}
where:
\begin{equation}\label{eq2.20}
\aleph(\alpha) = \frac{1}{2 \alpha} \left[
 Li_{\frac{3}{2}}(-\alpha) - Li_{\frac{5}{2}}(-\alpha) \right]
\end{equation}
$\aleph(\alpha)$ is  zero for $\alpha  = 0$, has  a maximum of  $0.0908$ at
 $\alpha = 7.8$, and goes to zero for $\alpha \rightarrow
 \infty$, which implies that for any $\epsilon^*$, the RPA 
becomes thermodynamically consistent in the high density limit.

Interestingly, within the RPA2,  the compressibility e.o.s.\ may also
be solved analytically and yields exactly the same result
(\ref{eq2.19}) as the virial e.o.s.\ in the RPA (i.e.\
$Z_{RPA2}^c=Z_{RPA}^v$), suggesting that the latter is more accurate
than the MF or RPA compressibility equation of state (\ref{eq2.14}).
In fact, as shown in Figures \ref{Fig2.4}-\ref{Fig2.6}, the RPA virial
equation of state is virtually indistinguishable from the practically
self-consistent HNC results and the MC simulations, except at very
low reduced densities $\rho^*$.  Fig.\ \ref{Fig2.6} demonstrates that
the mean-field approximation (\ref{eq2.14}) is still surprisingly
good, even for an interaction as large as $\epsilon^* =90$, just below
the value where freezing sets in.  This implies that the
hard-sphere limit, envisioned by Stillinger\cite{Stil76}, is still not
reached at such a strong interaction, and that the Gaussian core-model
behaves very much like a ``mean field fluid'' (MFF) over a wide
temperature and density range.

Perhaps the most striking result is the persistence of the linear
slope of the equation of state to such low densities\cite{lowdens}.  The slope
differs, however, from that determined by the (smaller) second virial
coefficient.  This is further discussed in Appendix B, where it is
shown that the standard virial expansion of the equation of
state\cite{Hans86} has a very small radius of convergence for the GCM,
and is of limited use for this model, contrarily to the case of the
hard sphere fluid\cite{Hoov67}.

\section{The GCM near a  wall}

In view of the success of the HNC and RPA theories for the GCM in the
homogeneous bulk fluid phase, it is of interest to ascertain the
validity
of these approximations under inhomogeneous conditions, e.g.\ in the
presence of an external potential $\phi({\bf r})$ acting on the
particles. In this section we shall consider more specifically the
density profiles of  GCM particles near a hard wall, using the
formalism of density functional theory (DFT).  In an external potential
the density of particles will change from a constant bulk value,
$\rho_b$, to a spatially varying local density $\rho({\bf r})$.  The
grand potential of the inhomogeneous fluid in equilibrium with a bulk
reservoir fixing the chemical potential $\mu$, may be cast in the
generic
form\cite{Evan92}
\begin{equation}\label{eq3.1}
\beta \Omega_v[\rho({\bf r})] = \beta {\cal F}^{in}[\rho({\bf r})] - 
\int d{\bf r}\left(\beta \mu - \beta \phi(r)\right) \rho({\bf r})
\end{equation}
where the intrinsic free energy functional ${\cal F}^{in}$ naturally splits 
into ideal and excess parts, ${\cal F}^{id}$ and ${\cal F}^{ex}$.  The
latter is an unknown functional of the local density $\rho({\bf r})$.
When the inhomogeneity is not too strong, the excess part ${\cal
F}^{ex}$ may be expanded in a functional Taylor series in the
deviation of the local density $\rho({\bf r})$ from the bulk density
$\rho_b$.  If the expansion is truncated after second order the HNC
functional results:
\begin{eqnarray}\label{eq3.2} 
\beta \Omega_v[\rho({\bf r})]   =  \beta  \Omega[\rho_b] + \int d {\bf
r'} \beta \phi({\bf r'}) \rho({\bf r'}) \nonumber \\  
+ \int d{\bf r'} \left\{ \rho({\bf r'})
 \ln \left[\rho({\bf r'})/\rho_b\right] - \rho({\bf r'})
 + \rho_b \right\}  \nonumber \\
  -  
\frac{1}{2} \int d{\bf r} d{\bf r'} 
(\rho({\bf r}) - \rho_b )c^{(2)}_b(|{\bf r} - {\bf r'}|) (\rho({\bf r'}) - \rho_b ).
\end{eqnarray}
This functional is to be minimized with respect to $\rho({\bf r})$,
and the resulting Euler-Lagrange equation reads:
\begin{eqnarray}\label{eq3.3}
\rho({\bf r}) &=& \rho_b \exp \Biggl[ -\beta \phi({\bf r}) \nonumber
\\
& +&  \rho_b \int d {\bf r'}  c^{(2)}_b(|{\bf r'}-{\bf r}|) \left(\frac{ \rho({\bf r'})}{\rho_b}-1 \right)
\Biggl]
\end{eqnarray}
which is the familiar HNC approximation for the density profile
$\rho({\bf r})$ in terms of the external potential and the {\em bulk} 
direct correlation function $c_b^{(2)}(r)\equiv c(r)$.
  Given $c(r)$ from the previous HNC
calculations
of the pair structure in the bulk, Eq.\ (\ref{eq3.3}) may be solved
iteratively for any $\phi({\bf r})$.  If $c(r)$ is replaced by its RPA 
form (\ref{eq2.4}), then Eq.\ (\ref{eq3.3}) reduces to the MF form:
\begin{eqnarray}\label{eq3.5}
\rho({\bf r}) &=& \rho_b \exp \Biggl[ -\beta \phi({\bf r}) \nonumber \\
 &-& \rho_b \int d{\bf r'}
\beta v(|{\bf r} - {\bf r'}|)\left( \frac{\rho({\bf r'})}{\rho_b}-1 \right)
 \Biggl].
\end{eqnarray}
Eq.\ (\ref{eq3.5}) also follows directly from the standard mean field
approximation (MF-DFT) for for
the intrinsic free energy functional\cite{Evan92}:
\begin{equation}\label{eq3.6}
\beta {\cal F}^{in}[\rho({\bf r})] = \beta {\cal F}^{id} + \frac{1}{2} \int
d{\bf r} d{\bf r'}  \beta  v({\bf r},{\bf r'}) 
\rho({\bf r}) \rho ({\bf r'}),
\end{equation}
which, in a different context, 
 is identical to the functional used to derive the Poisson-Boltzmann
 theory for ionic fluids if $v({\bf r},{\bf r'})$ is taken to be the
Coulomb potential and $\rho$ the charge density. 

Specializing to the case of a planar wall coinciding with the $x-y$
plane, and confining the particles to the $z \geq 0$ half-space, without
any additional external potential, we note that the density profile
$\rho(z)$ satisfies the contact condition\cite{Hend79}:
\begin{equation}\label{eq3.7}
\rho(z$$=$$0) = \beta P
\end{equation}
where $P$ is the pressure exerted by the particles on the wall, equal
to the bulk pressure in the absence of an external potential.  The sum
rule (\ref{eq3.7}) is satisfied by the MF-DFT approach, where
$\rho_{MF}(0)$$=$$\beta P_{MF}$$=$$\rho_b Z_{MF}$, with $Z_{MF}$ defined by
Eq.\ (\ref{eq2.14}) since $Z_{MF}$$=$$Z^c_{RPA}$.  However, the sum rule
 is not satisfied
by the (more accurate) HNC approximation (\ref{eq3.3}), which instead
leads to\cite{Carn81}:
\begin{equation}\label{eq3.8}
\rho(0) = \frac{1}{2} \rho_b \left[ 1 + \left(\frac{\partial \beta
P}{\partial \rho_b} \right)_T \right]
\end{equation}
\begin{figure}
\begin{center}
\epsfig{figure=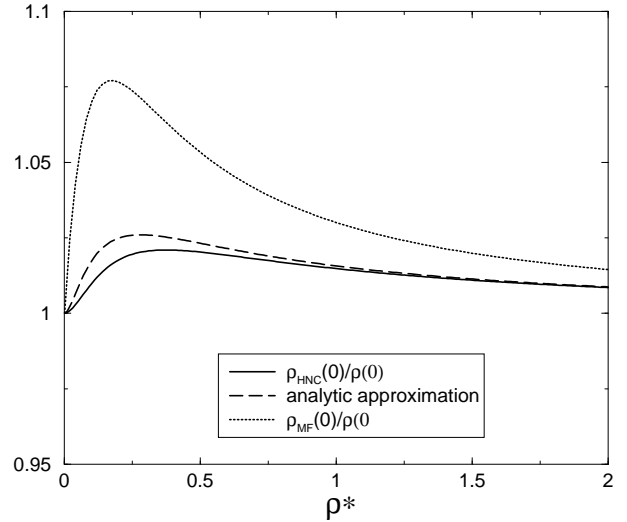,width=8cm} 
\begin{minipage}{8cm}
\caption{\label{Fig3.1}
Ratio $\rho_{HNC}(0)/\rho(0)$ from (\protect\ref{eq3.8}) compared to
the analytic  ratio obtained from the RPA2 approximation  to 
Eq.\ (\protect\ref{eq3.8}))  for $\epsilon^* =2$.
Also included is the ratio $\rho_{MF}(0)/\rho(0) \simeq
Z^c_{RPA}/Z^v_{HNC}$.  Clearly the HNC is the better approximation,
even
though it does not satisfy the sum-rule of Eq.\ (\protect\ref{eq3.7})
}
\end{minipage}
\end{center}
\end{figure}
This reduces to the exact result (\ref{eq3.7}) provided the pressure
is a quadratic function of the bulk density.  This is very nearly true
over a wide range of densities, as shown in the previous section.  In
particular the simple RPA compressibility (or MF) e.o.s.\ (\ref{eq2.14a}), which provides a fair representation of the numerical
HNC results, is of the necessary linear form to make Eq.\
(\ref{eq3.7}) and Eq.\ (\ref{eq3.8}) compatible.  The deviations from
the sum-rule (\ref{eq3.7}) may be traced back to the slight
non-linearity of $Z_{HNC}$, as demonstrated in the inset of Fig.\
\ref{Fig2.4} and in Fig.\ \ref{Fig3.1}. The relative error does not exceed $3\%$ for
$\epsilon^*=2$, although it tends to increase with increasing
$\epsilon^*$.  In fact, the ratio $\rho_{HNC}(0)/\rho(0)$ may be
estimated from the very accurate RPA2; since $Z^c_{RPA2} =
Z^v_{RPA}$, the required pressure may be calculated from
Eq.\ (\ref{eq2.19}), while the RPA2 inverse
compressibility is calculated to be :
\begin{equation}\label{eq3.9}
\left( \frac{\partial \beta P}{\partial \rho} \right)_T = 
1 + \alpha - \frac{\epsilon^*}{2 \alpha} \left[Li_{1/2}(-\alpha) -
Li_{3/2}(-\alpha) \right].
\end{equation}
The resulting analytic estimate of $\rho_{HNC}(0)/\rho(0)$ is also
shown in Fig.\ \ref{Fig3.1}; as expected, it gives a good
approximation of $\rho_{HNC}(0)/\rho(0)$.  Even though the MF approach
exactly satisfies the sum-rule (\ref{eq3.7}), the HNC approach, which
does not satisfy (\ref{eq3.7}), is a better approximation.  We note
that the arguments above can be extended to the popular Percus-Yevick
approximation\cite{Hans86}, where $\rho_{PY}(0)/\rho(0) \approx (1 +
1/2 \alpha)^(-1/2)$\cite{Carn81}, which, in contrast to the HNC or MF
approaches, becomes increasingly less accurate as the density increases.

\begin{figure}
\begin{center}
\epsfig{figure=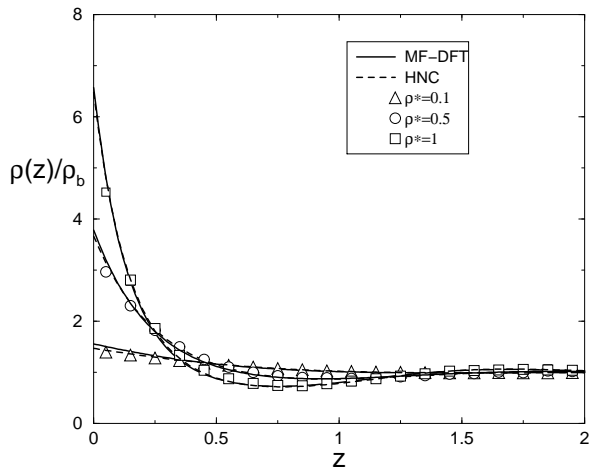,width=8cm} 
\begin{minipage}{8cm}
\caption{\label{Fig3.2} Density profiles from HNC and the MF-DFT for
Gaussian particles ($\epsilon = 2, R=1$) near a hard wall.  Symbols
are for MC simulations at 3 densities, the solid lines are from the
MF-DFT approach, and the dashed lines are from the HNC approach.  The
two theories and simulation agree to within the accuracy of the graph
for $\rho* = 0.5$ and $\rho* = 1$, but small discrepancies appear for
$\rho* = 0.1$, where the HNC is slightly more accurate.  }
\end{minipage}
\end{center}
\end{figure}

\begin{figure}
\begin{center}
\epsfig{figure=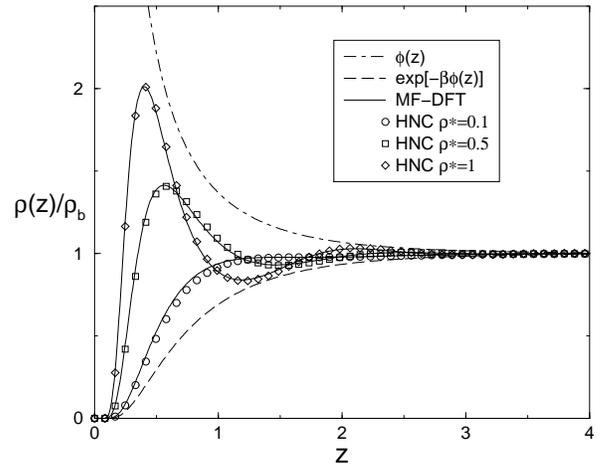,width=8cm} 
\begin{minipage}{8cm}
\caption{\label{Fig3.2b} Density profiles from HNC (symbols) and 
MF-DFT (lines) for Gaussian particles ($\epsilon = 2, R=1$)
interacting with an external potential $\beta \phi(z) = \exp[-z]/z$.}
\end{minipage}
\end{center}
\end{figure}

We have numerically solved  the HNC  and MF 
Euler-Lagrange  equations (\ref{eq3.3}) and (\ref{eq3.5})
 to calculate the density profiles $\rho(z)$ of a GCM near a hard
wall, for several values of the bulk density$\rho_b$.  The theoretical
profiles are compared in Fig.\ \ref{Fig3.2} to the results of MC
simulations.  The agreement is seen to be excellent, particularly at
the higher densities.  In fact, within the accuracy of the figure
 the difference between the HNC and
MF approaches is visible only for small $z$ at $\rho = 0.1$, where, as
expected, the HNC approach is slightly more accurate.
\vglue - 0.5cm In Fig.\ \ref{Fig3.2b} we show density profiles for
particles interacting with an external potential $\beta \phi(z) =
\exp[-z]/z$, a situation similar to that encountered for polymer coils
near a wall\cite{Loui00a}.  Once again, we observe that the HNC and
MF-DFT approaches are very close, implying that both are very accurate
and could potentially be fruitfully combined with the effective
potentials between polymer CMs\cite{Loui00a} to derive a full DFT
for polymer solutions in complex geometries.

In summary then, the results of this section confirm that the model
considered indeed behaves as a ``mean field fluid'' under
inhomogeneous conditions.
\vglue - 0.5cm

\section{Phase separation in two-component repulsive Gaussian
mixtures}
Since the underlying polymer mixtures
 exhibit interesting phase behavior
under a variety of physical conditions, it is natural to consider
binary mixtures of Gaussian core mixtures, interacting via pair
potentials:
\begin{equation}\label{eq8.1}
\beta v_{\nu \mu}(r) = \epsilon^*_{\nu \mu} \exp[-(r/R_{\nu \mu})^2],
\end{equation} 
where the species indices $1 \leq \nu, \mu \leq 2$.  The total 
number density is still denoted by $\rho = (N_1  + N_2)/V$, while the 
concentration variable $x = N_2/N$.  We are interested in the
possibility 
of a  phase separation, or de-mixing transition, of the two species.
The thermodynamic stability conditions for any binary mixture can be
expressed in terms of the Helmholtz free energy per particle, $f(x, v)
= F(N_1, N_2, V)/N$, considered as a function of the intensive
variables $x$ and $v$ (or $\rho = 1/v$), for any fixed temperature.
These conditions are\cite{Rowl82}:
\begin{mathletters}
\begin{eqnarray}
\left(\frac{\partial^2 f}{\partial v^2} \right)_x    & > 0
\label{eqC.1a}  \\
\left( \frac{\partial^2 f}{\partial x^2} \right)_v    & > 0
\label{eqC.1b} \\
\left(\frac{\partial^2 f}{\partial v^2} \right)_x \left(\frac{\partial^2 f}{\partial x^2} \right)_v  -
\left(\frac{\partial^2 f}{\partial v \partial x} \right)^2
    & > 0.
\label{eqC.1c} 
\end{eqnarray}
\end{mathletters}
The first inequality expresses mechanical stability, (i.e. positive
compressibility), the second is the condition for stability against
spontaneous de-mixing at constant volume while the last inequality
ensures stability at constant pressure; it is equivalent to the more
familiar condition $(\partial^2 g(x,P)/\partial x^2)_P > 0$, where $g$
is the Gibbs free energy per particle.  We note that if either of the
first two stability conditions (\ref{eqC.1a}) or (\ref{eqC.1b}) are
violated, the more restrictive stability condition, (\ref{eqC.1c}), is
violated as well.  Spinodal instability occurs when Eq.\
(\ref{eqC.1c}), is satisfied as an equality.  The condition is
equivalent to the $k \rightarrow 0$ divergence of the
concentration-concentration structure factor:
\begin{equation}\label{eqC.2}
S_{cc}(k) = x^2 S_{11}(k) + (1-x)^2 S_{22}(k) - 2x(1-x)S_{12}(k),
\end{equation}
where the $S_{\nu \mu}(k)$ are the usual partial structure
factors\cite{Hans86}.  From the OZ relations for a binary mixture it
is easily inferred that $S_{cc}(0)$ diverges when:
\begin{eqnarray}\label{eqC.3}
 \left[1 - (1-x)\rho \hat{c}_{11}(0)\right]
\left[1 - x\rho \hat{c}_{22}(0)\right]  \nonumber \\
 - x(1-x)\rho^2 \left[\hat{c}_{12}(0)\right]^2 = 0.
\end{eqnarray}

We now examine the implications of these conditions within the MF
approximation, which we have shown to yield reliable results, except
at low reduced density $\rho^*$.  The MF free energy (\ref{eq2.15}),
properly generalized to the binary situation, reads:
\begin{equation}\label{eqC.4}
f(x,\rho) = f^{id}(\rho) + f^{mix}(x) + \frac{1}{2} \rho \hat{V}_0(x)
\end{equation}
where the first, ideal gas, term is irrelevant in the subsequent
considerations, $f^{mix}$ is the ideal mixing term:
\begin{equation}\label{eqC.5}
f^{mix}(x) = x \ln x + (1-x) \ln (1-x),
\end{equation}
and the MF interaction term is:
\begin{equation}\label{eqC.6}
\hat{V}_0(x) = (1-x)^2 \beta \hat{v}_{11}(0) + 2x(1-x)\beta \hat{v}_{12}(0) + 
x^2\beta\hat{v}_{22}(0).
\end{equation}
The $\{\beta \hat{v}_{\nu \mu}(0)\}$ are the $k\rightarrow 0$ limits of the
FTs of the interaction potentials.   In fact, the MF free-energy
(\ref{eqC.4}) has the same mathematical form as a second-virial theory
which would be valid for very low densities\cite{vanR96}.

With the MF free energy (\ref{eqC.4}), the stability conditions
(\ref{eqC.1a} - \ref{eqC.1c}) reduce to:
\begin{mathletters}
\begin{eqnarray}
1 + \rho \hat{V}_0(x)
 & > 0
\label{eqC.8a}  \\
1 - \rho x (1-x) \chi
 & > 0
\label{eqC.8b} \\
1 + \rho \hat{V}_1(x) - \rho^2 x(1-x) \Delta
    & > 0
\label{eqC.8c} 
\end{eqnarray}
\end{mathletters}
respectively, 
where the following parameters were defined:
\begin{eqnarray}
\chi  &=&  2\beta\hat{v}_{12}(0) -(\beta\hat{v}_{11}(0) +\beta\hat{v}_{22}(0)) 
\label{eqC.9a}\\
\Delta &=& (\beta\hat{v}_{12}(0))^2 - \beta\hat{v}_{11}(0)
 \beta\hat{v}_{22}(0)\label{eqC.9b} \\
 \hat{V}_1(x) &=&
 (1-x)\beta\hat{v}_{11}(0) + x \beta \hat{v}_{22}(0) \label{eqC.9c}
\end{eqnarray}
Eq.\ (\ref{eqC.8a}) can only be violated if the potentials themselves
 violate a 2-component extension of Eq.\ (\ref{eqA.3}) from Appendix A,
 which is a necessary (but not sufficient) condition for the existence
 of a well-defined thermodynamic limit.  The limit of stability of the
 mixture (i.e. the spinodal line) at constant volume or pressure is
 reached when the inequalities (\ref{eqC.8b}) and (\ref{eqC.8c}) turn into
 equalities; the latter condition also follows from (\ref{eqC.3}), when
 the $\hat{c}_{\nu \mu}$ are replaced by their RPA limits
 $\hat{c}_{\nu \mu}(k) = - \beta \hat{v}_{\nu \mu}(k)$ .  De-mixing at
 constant volume is possible, provided $\chi > 0$; the density along
 the spinodal is then easily calculated to be:
\begin{equation}\label{eqC.10}
\rho_s(x) = \frac{1}{x(1-x)\chi}.
\end{equation}
De-mixing at constant pressure is only possible provided $\Delta > 0$.
The corresponding density along the spinodal satisfies:
\begin{equation}\label{eqC.11}
\rho_s(x) = \frac{\hat{V}_1(x) 
+ \sqrt{\hat{V}_1(x)^2 + 4x(1-x)\Delta}}{2x(1-x) \Delta},
\end{equation}
the pressure along the spinodal is:
\begin{equation}\label{eqC.12}
P_s(x) = \rho_s(x) + \frac{1}{2} \rho_s^2(x) \hat{V}_0(x),
\end{equation}
and the critical consolute point is determined by the condition:
\begin{equation}\label{eqC.13}
\frac{d P_s(x)}{d x} = 0.
\end{equation}
The simple quadratic expression for the pressure P (\ref{eqC.12}), is easily inverted to
obtain
an expression for the spinodal density as a function of concentration
$x$ and pressure $P$.

We now apply these general considerations within the MF framework to
the binary Gaussian core model for which 
\begin{equation}\label{eqC.7}
\beta \hat{v}_{\nu \mu}(0) = \pi^{3/2} \epsilon^*_{\nu \mu} R_{\nu \mu}^3.
\end{equation} Inserting the binary GCM expression
for $\hat{v}_{\nu \mu}(0)$ into  expressions (\ref{eqC.9b}) and
(\ref{eqC.9c}) for $\chi$ and $\Delta$, we find that phase-separation
at constant volume or at constant pressure is possible provided:
\begin{equation}\label{eqC.14}
\chi = \pi^{3/2} \left[2 \epsilon^*_{12} R_{12}^3 - 
( \epsilon^*_{11} R_{11}^3 +\epsilon^*_{22} R_{22}^3 )\right] > 0.
\end{equation}
or:
\begin{equation}\label{eqC.15}
\Delta = \pi^3\left[(\epsilon^*_{12})^2 R_{12}^6 - 
 \epsilon^*_{11}   \epsilon^*_{22} R_{11}^3 R_{22}^3 )\right] > 0.
\end{equation}

In order to focus on physically relevant values of the parameters
$\epsilon^*_{\nu \mu}$ and $R_{\nu \mu}$,  it is important to make
contact 
with known results for polymer coils in a good
solvent\cite{Krug89,Daut94}.  Simulations on binary solutions of
self-avoiding polymer coils carried out in the low concentration
limit\cite{Daut94} suggest that the effective pair potentials between
centers of mass are reasonably well represented by the Gaussian form
(\ref{eq8.1}), with :
\begin{equation}\label{eqC.16}
\epsilon^*_{12} \leq \epsilon^*_{11} \simeq \epsilon^*_{22}
\end{equation}
and 
\begin{equation}\label{eqC.17}
R_{12}^2 \simeq \frac{1}{2}\left(R_{11}^2 + R_{22}^2\right).
\end{equation}
The relation (\ref{eqC.16}) between the $\epsilon_{\nu \mu}$ 
favors mixing.  On the other hand
$R_{12} > 1/2(R_{11} + R_{22})$, which resembles the 
positive non-additivity that can drive de-mixing
in hard-core mixtures\cite{Loui00}.
Substituting (\ref{eqC.17}) into (\ref{eqC.14}), we find that
a spinodal instability of the mixture is possible at constant volume
provided:
\begin{equation}
\frac{\epsilon^*_{12}}{\epsilon^*_{11}} > \sqrt{2} \frac{1 + (R_{22}/R_{11})^3}{\left(1
+ (R_{22}/R_{11})^2\right)^{3/2}} \geq 1
\end{equation}
which contradicts the requirement (\ref{eqC.16}).  On the other hand,
if (\ref{eqC.17}) is substituted into (\ref{eqC.15}), de-mixing at
constant pressure may occur provided:
\begin{equation}\label{eqC.18}
\frac{(\epsilon_{12}^*)^2}{\epsilon^*_{11} \epsilon^*_{22}} >
 \left[ \frac{2 (R_{22}/R_{11})}{1 + (R_{22}/R_{11})^2} \right]^3 \leq 1,
\end{equation}
which is compatible with the requirement (\ref{eqC.16}).

\begin{figure}
\begin{center}
\epsfig{figure=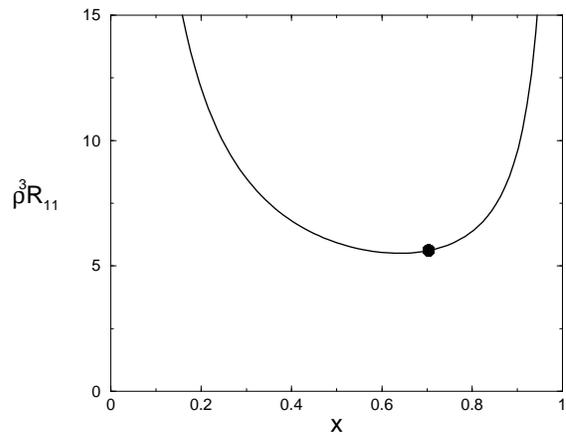,width=7cm} 
\begin{minipage}{8cm}
\caption{\label{FigC.1} Constant pressure spinodal
(\protect\ref{eqC.11}) for parameters taken from simulations of
$L=100$ and $L=200$ monomer effective polymer CM
potentials\protect\cite{Daut94}.  The x-axis denotes the composition
$x = x_2= N_2/N$.  The y-axis denotes the density $\rho R_{11}^3$,
where $R_{11}$ is the = radius of gyration $R_G$ for the $L=200$
polymers.  The dot is the critical point at $(x=0.70,\rho R_{11}^3=5.6)$. }
\end{minipage}
\end{center}
\end{figure}

More specifically, we have chosen values of the parameters
$\epsilon^*_{\nu \mu}$ and $R_{\nu \mu}$ appropriate for a polymer
mixture of self-avoiding polymers of $L=200$ (species $1$) and $L=100$
(species 2) monomers\cite{Daut94,Watz99}.  The resulting spinodal line
(\ref{eqC.11}) in the $x-\rho$ plane, calculated from the MF free
energy (\ref{eqC.4}) with (\ref{eqC.7}), is shown in Fig.\
\ref{FigC.1}.  Phase separation into two solutions of different
composition $x$ occurs above a critical density $\rho^*_c =
5.6/R_{11}^3$ and critical composition $x_c = 0.70$.  Note that since
all terms in the free energy are of entropic origin, the temperature
scales out, i.e.\ the mixture behaves as an athermal system.  In view
of the remarkable accuracy of the MF theory at high density, as
illustrated in sections II and III for the one component GCM fluid, we
expect the phase-diagrams, calculated within MF (or equivalently RPA)
to be reliable; full calculations of the binodal line, based on RPA2
and HNC theories, will be reported elsewhere.
\vglue - 0.5cm 

\section{conclusions}
The calculations carried out in this paper, and in related
work\cite{Loui00a}, lead to the conclusion that a system of classical
particles interacting via a repulsive Gaussian core potential behaves
like a weakly correlated ``mean field fluid'' over a wide range of
temperatures and densities.  In fact for any temperature there is
always a (surprisingly low) density beyond which the excess free
energy per particle is a linear function of density and the resulting
excess pressure increases like a quadratic function of the density. On
the other hand, in the opposite low density regime, a virial expansion
of the equation of state in powers of the density appears to converge
only at extremely low densities.  This is in sharp contrast to
hard-core systems, for which the virial expansion provides a good
estimate of the equation of state up to relatively high packing
fractions\cite{Hoov67}, while the pressure diverges near close packing
according to a simple free volume picture.  At very strong interaction
strength $(\epsilon^* \gtrsim 100)$, the GCM behaves effectively as a
hard core fluid that freezes at intermediate densities, but re-melts
under further compression to return to mean-field like
behavior\cite{Stil76,Lang00}. The small correlational effects at low
and intermediate densities are adequately described by the simple,
analytic RPA2 extension of RPA theory, or by the HNC integral equation
(requiring numerical solution), which is nearly thermodynamically
consistent over a broad range of temperatures and densities.

The MF theory performs equally well in the inhomogeneous situation of
Gaussian core particles near a hard wall.  The binary version of the
model phase-separates at high densities, when the widths of the
Gaussian repulsion satisfy the composition rule (\ref{eqC.17}), provided
condition (\ref{eqC.18}) is satisfied.  This provides an interesting example of
phase separation in systems with purely repulsive interactions.  

To conclude it seems worthwhile to consider the relevance of the GCM
for the description of polymer solutions.  The latter enter the
semi-dilute regime when polymer coils start to overlap, i.e. when
$\rho* \sim 3/(4\pi)$. For densities of this order we have seen that
the GCM behaves like a ``mean field fluid'', with a quadratic density
dependence of the pressure.  The exponent $2$ is close to the $9/4$
power observed for the osmotic pressure of semi-dilute polymer
solutions\cite{Doi95}.  The difference between the exponent $9/4$ and
$2$ is due in part to the weak, but significant density dependence of
the effective pair potential between the centers of mass of
self-avoiding polymers\cite{Loui00a}, which leads to an additional
density dependence of the RPA or MF equation of state (\ref{eq2.14}).
This possibility is being explored in more detail\cite{Bolh00a}.

The effective polymer-wall potentials derived in ref. \cite{Loui00a}
show a significant variation with density.  Nevertheless, the form of
the $\rho(z)/\rho_b$ for the GCM in a {\em fixed} external potential
follows the same qualitative trends as the distribution of the polymer
CM's near a wall, $\rho_{CM}(z)/\rho_b$, suggesting that the physics
of polymer coils near a wall is well captured by the GCM.

The de-mixing transition of binary Gaussian core mixtures is
reminiscent of the tendency of polymers of different molecular weight
to phase separate at high concentration and in the melt. 
 Again,
further analysis is required to decide if the analogy between the
de-mixing of Gaussian core mixtures and of polymer blends is
fortuitous, or has some deeper foundation.

\acknowledgements 

AAL acknowledges support from the Isaac Newton Trust, Cambridge, PB
acknowledges support from the EPSRC under grant number GR$/$M88839,

We thank Christos Likos for helpful correspondence, Bob Evans for
suggesting the mean-field DFT approach to us, and David Rowan for a
critical reading of the manuscript.

\appendix

\section{Thermodynamic stability of soft-core potential systems}\label{App1}

It was pointed out in section \ref{secIIa} that the GCM satisfies Ruelle's
condition (\ref{eq2.2}) for the existence of a finite thermodynamic
limit.  In this appendix we give two examples of pair potentials
involving a repulsive core and a small {\em attractive} component which
do not satisfy Ruelle's stability condition, and hence belong to the
class of potentials referred to by him as ``catastrophic''.  The
following considerations are not completely academic, since it has been
shown in ref \cite{Bolh00a} that the effective pair potential between
the centers of mass of two polymer coils in a good solvent indeed
exhibits a small attractive part at distances of the order of several
times the radius of gyration $R_g$ for intermediate densities. When
the polymer coils are no longer in a good solvent the potentials can 
develop even larger attractive parts\cite{Daut94}.

According to proposition $3.2.2$ in Ruelle\cite{Ruel69}, given an
interaction energy $V_N^{(2)}$ built up by pair potentials $v_2$, the
grand partition function is finite only if the following two
equivalent properties hold for all 
$N \geq 0$ and all $\{{\bf r_i}\} \in {\bf R^N}$:
\begin{equation}\label{eqA.1a}
\sum_i^N \sum_j^N v_2 (|{\bf r_i - r_j}|) \geq 0
\end{equation}
and
 \begin{equation}\label{eqA.1b}
V_N^{(2)}({\bf r_1},....,{\bf r_N})= \sum_{1 \leq i < j \leq N} v_2(|{\bf
r_i} - {\bf r_j}|)\geq -NB
\end{equation}
for a $B \geq 0$.
 Note that
in (\ref{eqA.1a}) the double sum includes the self-interaction
($i=j$).
\begin{figure}
\begin{center}
\epsfig{figure=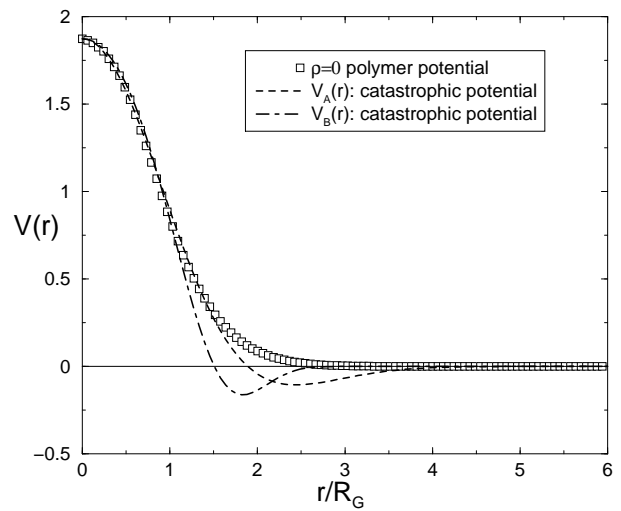,width=8cm} 
\begin{minipage}{8cm}
\caption{\label{FigA.1} Two ``catastrophic'' potentials compared to a
typical CM potential for two polymers in a good solvent.  Potential
$v_A(r)$ (\protect\ref{eqA.2a}) violates the conditions (\ref{eqA.1a}) and
(\ref{eqA.1b})  for homogeneous fluid configurations, while potential
$v_B(r)$ (\protect\ref{eqA.2b}) violates (\ref{eqA.1a}) and
(\ref{eqA.1b})   only
for inhomogeneous configurations like the fcc crystal.}
\end{minipage}
\end{center}
\end{figure}
\vglue - 0.5cm

We consider two examples of  potentials which do not
satisfy these conditions:
\begin{eqnarray}
v_{A}(r) &= & 1.87\cos\left[\sqrt{(2+\delta)}
\left(\frac{r}{1.7R_G}\right)\right]exp\left[-\left(\frac{r}{1.7R_G}\right)^2\right]
\label{eqA.2a}\\
v_{B}(r) &= & 1.87\cos\left[\sqrt{\pi} \left(\frac{r}{1.7R_G}\right)\right]exp\left[-\left(\frac{r}{1.7R_G}\right)^4\right] \label{eqA.2b},
\end{eqnarray}
and compare them in Fig.\ \ref{FigA.1} to the polymer CM potential
between two isolated $L=500$ SAW polymer coils. Here $\delta$ is an
arbitrary positive constant taken to be $\delta = 0.001$ in
Fig. \ref{FigA.1}. Although at first sight they don't appear very
different from the purely repulsive polymer potential, they are both
``catastrophic''.

If $\delta > 0$, the first potential, $v_A(r)$, violates a weaker
condition than (\ref{eqA.1a}) or (\ref{eqA.1b}), namely
\begin{equation}\label{eqA.3}
\hat{v}(0)= \int v(r) d{\bf r} > 0,
\end{equation}
which is necessary (but not sufficient) for a thermodynamic limit.
When it doesn't hold, conditions (\ref{eqA.1a}) and (\ref{eqA.1b}) can
be violated for a homogeneous ``gas'' with $g(r)=1$.  This has a
further implication for fluids described by a mean-field free-energy
(\ref{eq2.15}), since the inverse compressibility $\partial \beta
P/\partial \rho = 1 + \beta \hat{v}(0) \rho$ cannot go through zero
without violating the condition (\ref{eqA.3}), which implies that
one-component soft-core fluids described by a mean-field e.o.s.\
cannot support a spinodal instability.

The second potential, $v_B(r)$, has an integral $\hat{v}_B(0) > 0$, but
it still violates (\ref{eqA.1a}) and (\ref{eqA.1b}) for an {\em
inhomogeneous} configuration. For example, for an fcc lattice with
single occupancy $ \sum_i^N \sum_j^N v_B (|{\bf r_j} - {\bf r_i}|) =
-0.13N$.  The potential is catastrophic because one can always lower
the total energy indefinitely through multiple occupancy of the
lattice sites.

The $\theta$ point in polymer solutions can be defined as the temperature
where the effective second osmotic virial coefficient, $B_2$, passes through
$0$\cite{Daut94}.  Above the $\theta$ point the solvent is said to be
``good'', while below the $\theta$ point the solvent is said to be
``poor''.  Simulations of a model for two polymers in a poor solvent
show that the effective pair potential is no longer strictly positive
definite below the $\theta$ point\cite{Daut94}, implying that the pair
potentials can become catastrophic.  In fact, for the type of polymer
CM potentials considered, this seems to occur just below the $\theta$ point
temperature where $B_2=0$. It is tempting to speculate that the coil-globule
transition, which also typically occurs slightly below the $\theta$
temperature, is related to the point at which the effective pair
potential becomes catastrophic.  However, it is not yet clear whether
the pair-potential picture of polymer solutions\cite{Loui00a} remains
valid for poor solvents.

\section{Virial expansion for the GCM fluid}\label{App2}

In this appendix we briefly consider the convergence of the virial
expansion of the equation of state of the GCM in powers of the density
$\rho$.  The FT of the Mayer f-function in Eq.\ (\ref{eq2.8}) is given
by the convergent sum:
\begin{equation}\label{eqB.1}
\hat{f}(k) = \pi^{3/2}\sum_{n=1}^{\infty}
\frac{\exp(-\frac{k^2}{4n})(-\epsilon^*)^n}{n! n^{3/2}}.
\end{equation}
Here  the width parameter $R$ in the Gaussian potential (\ref{eq1.1})
has been chosen as unit of length for convenience.
The second and third virial coefficients $B_2$ and $B_3$, of the GCM
can then be expressed as  the following convergent sums:
\begin{eqnarray}\label{eqB.2}
B_2 &=& - \frac{1}{2} \hat{f}(0) = -\frac{\pi^{3/2}}{2} \sum_{n=1}^{\infty}
\frac{(-\epsilon^*)^n}{n! n^{3/2}} \\
B_3 &=& -\frac{1}{3} \pi^3 \sum_{i=1}^{\infty}\sum_{j=1}^{\infty}\sum_{k=1}^{\infty}
\frac{(-\epsilon^*)^{i+j+k}}{i!j!k!(ij+jk+ik)^{3/2}}\label{eqB.3}
\end{eqnarray}
\begin{figure}
\begin{center}
\epsfig{figure=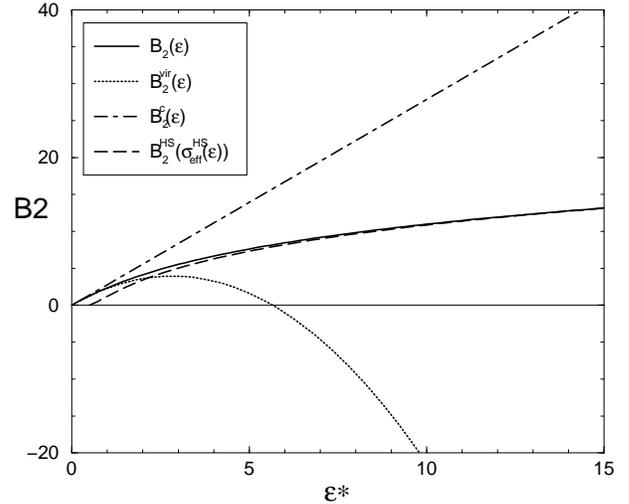,width=8cm} 
\begin{minipage}{8cm}
\caption{\label{FigB.1} Second virial coefficients for a Gaussian
potential as a function of interaction strength $ \epsilon^*$.
$(B_2^c > B_2 > B_2^{vir})$.  Also included is the empirical relation
$B_2^{HS}(\sqrt{\ln(2 \epsilon^*)})$, where
$B_2^{HS}(\sigma)$ is the hard-sphere second-virial coefficient.  }
\end{minipage}
\end{center}
\end{figure}
\vglue -0.5cm
The variations of $B_2$ and $B_3$ with $\epsilon^*$ are shown in Figs
\ref{FigB.1} and \ref{FigB.2}; both virial coefficients are always
positive.

 The virial expansion of the equation of state reads:
\begin{equation}\label{eqB.5}
Z = \frac{\beta P}{\rho} = 1 + B_2 \rho + B_3 \rho^2 + {\cal O}(\rho^{3})
\end{equation}
and the results from the $2$ and $3$ term series are compared in
Figs.\ \ref{Fig2.4}-\ref{Fig2.6} of section \ref{secII} to the predictions of the RPA and HNC
theories, and to MC simulations for $\epsilon^* = 2, 10$ and $90$.  The
virial expansion is seen to break down very early.  In particular,
although the MF e.o.s., which becomes very accurate at high
density, predicts a linear variation of $Z$ with density, the slope
differs more and more from $B_2$ as the interaction strength $\epsilon^*$ increases.  Adding the $B_3$ contribution leads to rapid
deterioration of the predicted e.o.s.\ as the density
increases.

 The shortcoming of the
virial expansion in powers of density is further illustrated by
considering the RPA.  From Eqns.\ (\ref{eq2.14a}) and (\ref{eq2.19}),
one may extract the following compressibility and virial estimates of
the 2nd and 3rd virial coefficients:
\begin{eqnarray}\label{eqB.6a}
B_2^{c} & =&  \frac{1}{2} \pi^{3/2}  \epsilon^*  ;  B_3^c =0 \\
B_2^{vir} & =&  \frac{1}{2} \pi^{3/2} \epsilon^* \left(1 -
 \frac{\sqrt{2}}{8} \epsilon^* \right) ; B_3^{vir} = \frac{\pi^3 (\epsilon^*)^3}{9 \sqrt{3}}\label{eqB.6b}
\end{eqnarray}
As shown in Figs.\ \ref{FigB.1} and \ref{FigB.2}, the exact virial
coefficients are bracketed by the virial and compressibility estimates
extracted from the RPA:
\begin{eqnarray}\label{eqB.7}
B_2^c > B_2 > B_2^{vir} \\ 
B_3^{vir} > B_3 > B_3^c =0
\end{eqnarray}
The large deviations shown in Figs.\ \ref{FigB.1} and
\ref{FigB.2} imply that, in contrast to the case at high densities,
the RPA is expected to perform poorly at very low densities and large
$\epsilon^*$, where it is thermodynamically inconsistent.  In fact,
$B_2^{vir}$ even goes negative for $ \epsilon^* \gtrsim 5.7$!  The
effect this has on the RPA virial e.o.s.\ is demonstrated in the inset
of Fig. \ref{Fig2.5} and in Fig. \ref{FigB.3}.  However, even though
for $\epsilon^* =2$, $B_2^{vir}$ is $13 \%$ less than $B_2$, and
$B_3^{vir}$ is over $300 \%$ larger than $B_3$, $Z_{RPA}^v$ remains  within
$1\%$ of the exact e.o.s.\ over the entire density range!  Thus, in
spite of the fact that the RPA virial approximation grossly
misrepresents the first two virial coefficients, it nevertheless
accurately describes the e.o.s., implying that the density is not a
good expansion parameter for the GCM fluid phase.
\begin{figure}
\begin{center}
\epsfig{figure=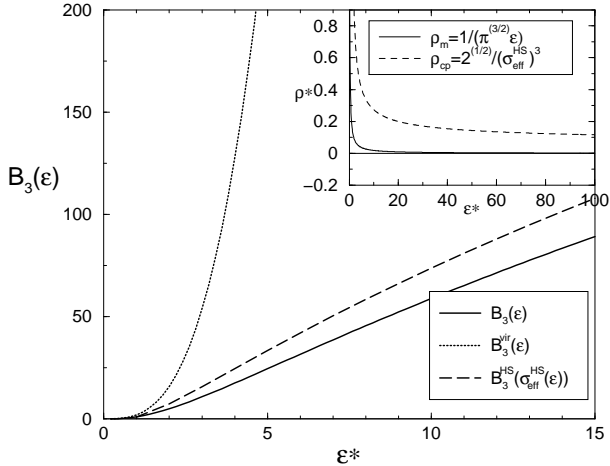,width=8cm} 
\begin{minipage}{8cm}
\caption{\label{FigB.2} Third virial coefficients for a Gaussian
potential as a function of interaction strength $\epsilon^*$.  $B_3^c
= 0$, $B_3^{vir} > B_3$. Also included is the empirical relation
$B_3(\epsilon^*) = B_3^{HS}(\sqrt{\ln(2 \epsilon^*)})$, where
$B_3^{HS}(\sigma)$ is the hard-sphere third-virial coefficient.
 {\bf Inset}: $\rho_m = 1/(\pi^{3/2}
\epsilon^*)$ is the maximum density for which the RPA virial e.o.s.\
(\ref{eq2.19}) can be written as an expansion in powers of the
density. $\rho_{cp} = \sqrt{2}/(\sigma^{HS}_{eff})^3$ is the density
at which the effective hard-sphere system with the same second virial
coefficient as the GCM would be close-packed, and beyond which a
virial equation would no longer be expected to exist.  }
\end{minipage}
\end{center}
\end{figure}
\vglue - 0.5cm

A further hint at the breakdown of the virial expansion comes from
summing the  virial series to all orders in the high temperature
limit, where, from the diagramattic representation of the
virial coefficients\cite{Hans86}, it can be shown that   the $B_n$ 
are given by:
\begin{equation}\label{eqB.4}
B_n = -\frac{1}{2} 
\pi^{\frac{3(n-1)}{2}} \frac{(-\epsilon^*)^n (n-1)}{n^{5/2}}
+ {\cal O}(\epsilon^*)^{n+1}
\end{equation}
for $n \geq 3$.  This, $B_2^{vir}$ (\ref{eqB.6b}) is added, 
recovers exactly the virial RPA equation of state (\ref{eq2.19}),
which can only be expanded in powers of density for $\alpha < 1$,
i.e.\ for
\begin{equation}\label{eqB.8}
\rho^* < \rho_m^* = \frac{1}{\pi^{3/2} \epsilon^*} \approx 0.1796/\epsilon^*.
\end{equation}
This implies that in the high-temperature limit, the virial expansion
does not converge for densities higher that $\rho_m^*$.  For
$\epsilon^* =2$, there is no convergent density expansion of the RPA
virial e.o.s.\ for $\rho* > \rho_m^* \approx 0.0898$.  A similar
breakdown in convergence may be expected for the exact virial
expansion.  The physical reason for this lack of convergence lies in
the possibility of multiple overlap of soft core particles, giving
much more weight to higher order cluster integrals compared to the
case of fluids with hard-core interactions.

At large enough $\epsilon^*$, the overlap probability becomes
exponentially small, and the GCM can be mapped onto an effective
hard-sphere system\cite{Stil76,Lang00}.  One possible criterion for
the mapping is to equate the second virial coefficients. From this we
obtain an effective hard-sphere radius of:
\begin{equation}\label{eqB.9}
\sigma^{HS}_{eff}(\epsilon) = \left(\frac{3}{2 \pi}
B_2(\epsilon^*)\right)^{1/3}
\end{equation}
which for $\epsilon^* > 1$ is well approximated by the empirical
expression $\sigma^{HS}_{eff} \approx \sqrt{\ln(2 \epsilon^*)}$. For
large $\epsilon^*$ and low densities, the equation of state resembles
that of hard-spheres (see e.g. Fig.\ref{FigB.3}), suggesting that a
virial expansion does indeed exist for low densities.  We note that
for this large value of $\epsilon^*$, the true virial expansion
appears to have a larger radius of convergence than that of the RPA
virial e.o.s., for which $\rho_m (\sigma^{HS}_{eff})^3 \approx 0.023$.
For $\epsilon^* \gtrsim 100$ there is a freezing transition at roughly
the density expected for the effective hard-sphere system ($\rho
(\sigma^{HS}_{eff})^3 \approx 1$), not far above which any effective
virial expansion is expected to break down (see e.g.\ the inset of
Fig.\ \ref{FigB.2}).  In fact, since Gaussian potentials don't have an
infinitely hard core, it is possible to achieve much higher densities
than are normally available to simple liquids.  At the lowest
densities the fluid is described by a linear second virial theory
e.o.s., but as the density increases, this rapidly turns over to a
mean field like linear e.o.s.\ with a different (larger) slope.  Thus,
even though the e.o.s.\ is well described by a first order polynomial
in the density $\rho$ it is not at all equivalent to a second virial
theory, and the density is generally not a good expansion parameter.

 \begin{figure}
\begin{center}
\epsfig{figure=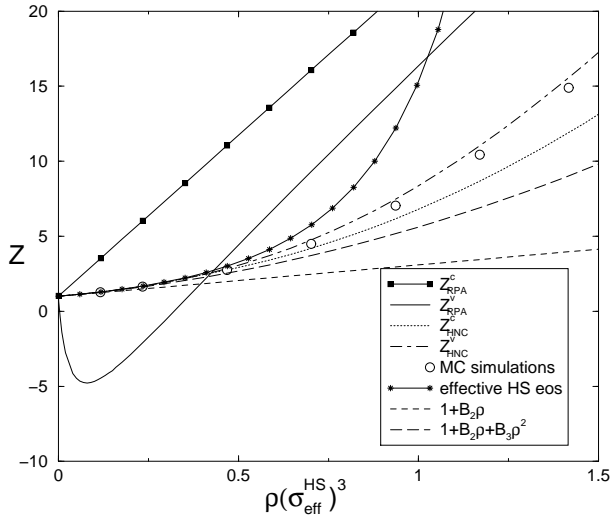,width=8cm} 
\begin{minipage}{8cm}
\caption{\label{FigB.3} $Z$ v.s.\ $\rho (\sigma^{HS}_{eff})^3$ in the
low density limit. Here $\epsilon^* =90$ so that $(\sigma^{HS}_{eff})
= 2.27$; $\rho (\sigma^{HS}_{eff})^3 = 1$ corresponds to $\rho^* =
0.085$.  For low effective density the e.o.s.\ follows the hard-sphere
e.o.s.\ (here approximated by the Carnahan-Starling form
\protect\cite{Hans86}). For higher densities the fluid moves towards
the mean-field fluid limit (see Fig. \ref{Fig2.6}). Note that the two
RPA expressions for the e.o.s.\ are very poor approximations in this
low density regime.  }
\end{minipage}
\end{center}
\end{figure}
\vglue - 0.5cm

\end{multicols}

\begin{thebibliography}{99}
\bibitem{polymer} We note that in the polymer literature the crossover 
density between the dilute and semi-dilute regime is often denoted by
a fixed density $\rho* \approx 3/(4\pi R_G^3)$.  However, in keeping
with traditional usage in the theory of simple
fluids\protect\cite{Hans86}, we prefer to use $\rho* = \rho R^3$ as a
reduced  density.
\bibitem{RPA}We prefer to use the term RPA instead of MSA, since
the latter usually implies a hard-core\cite{Hans86}.
\bibitem{lowdens} Of course at the very lowest densities the e.o.s.\
is again described by a linear second virial theory, but the slope is
smaller than that of the MF theory.
\end{thebibliography}
\end{document}